# Atomic Imaging of Mechanically Induced Topological Transition of Ferroelectric Vortices

Pan Chen[1,2,8,12], Xiangli Zhong[3,12], Jacob A. Zorn[4], Mingqiang Li[2,6], Yuanwei Sun[2], Adeel Y. Abid[2], Chuanlai Ren[3], Yuehui Li[6], Xiaomei Li[1], Xiumei Ma[2], Jinbin Wang[3], Kaihui Liu[7], Zhi Xu[1,8,9], Congbing Tan[5]★, Longqing Chen[4], Peng Gao[2,6,10,11]★, and Xuedong Bai[1,8,9]★

[1]*Beijing National Laboratory for Condensed Matter Physics, Institute of Physics, Chinese Academy of Sciences, Beijing 100190, China.*

[2]*Electron microscopy laboratory, School of Physics, Peking University, Beijing 100871, China.*

[3]*School of Materials Science and Engineering, Xiangtan University, Hunan Xiangtan 411105, China.*

[4]*Department of Materials Science and Engineering, Penn State University, University Park, Pennsylvania 16802, USA.*

[5]*Department of Physics and Electronic Science, Hunan University of Science and Technology, Hunan Xiangtan 411201, China.*

[6]*International Center for Quantum Materials, Peking University, Beijing 100871, China.*

[7]*State Key Laboratory for Artificial Microstructure & Mesoscopic Physics, School of Physics, Peking University, Beijing 100871, China.*

[8]*School of Physical Sciences, University of Chinese Academy of Sciences, Beijing 100190, China.*

[9]*Songshan Lake Materials Laboratory, Dongguan, Guangdong 523808, China.*

[10]*Collaborative Innovation Center of Quantum Matter, Beijing 100871, China*

[11]*Beijing Academy of Quantum Information Sciences, Beijing 100193, China*

[12]*These authors contributed equally: P. Chen and X. L. Zhong.*

★ E-mails: p-gao@pku.edu.cn; cbtan@xtu.edu.cn; xdbai@iphy.ac.cn

**Ferroelectric vortices formed through complex lattice–charge interactions have great potential in applications for future nanoelectronics[1-4] such as memories. For practical applications, it is crucial to manipulate these topological states under external stimuli. Here, we apply mechanical loads to locally manipulate the vortices in a PbTiO$_3$/SrTiO$_3$ superlattice via atomically resolved *in situ* scanning transmission electron microscopy. The vortices undergo a transition to the *a*-domain with in-plane polarization under external compressive stress and spontaneously recover after removal of the stress. We reveal the detailed transition process at the atomic scale and reproduce this numerically using phase-field simulations. These findings provide new pathways to control the exotic topological ferroelectric structures for future nanoelectronics and also valuable insights into understanding of lattice-charge interactions at nanoscale.**

Ferroelectric vortices composed of electric dipole moments with continuous rotation are theoretically predicted to occur in nanostructures such as nanowires[5], nanodots[6] and nanocomposites[7] and have been experimentally realized in rhombohedral $BiFeO_3$ [8-11], tetragonal $PbZr_{0.2}Ti_{0.8}O_3$ [12] thin films, and $PbTiO_3/SrTiO_3$ (PTO/STO) superlattices[3]. These polarization vortex states that are regarded as topological defects and can be characterized by an electric toroidal moment, can exist over a few nanometers, making them a perfect candidate for data storage applications[13], as their storage capacity can reach ~$10^{12}$ bits per square inch with little cross-talk between adjacent bits, which is several orders of magnitude larger than the current technology of ferroelectric storages[1].

To enable the application of the polarization vortices in the systems described above in nanotechnology, we need to be able to effectively manipulate the order parameters and phase transitions under external stimuli. Theoretical studies have proposed switching either by a curled electric field[14] or by rationally designed nanostructures, such as nanorings[15] or notched nanodots[16]. Experimental investigation was only limited in a phase coexistence (vortex and $a_1/a_2$ phase) system by Damodaran et al.[17], who used an atomic force microscopy (AFM) tip to apply an external electric field to a mixed phase system and demonstrated the interconversion between vortices and the regular ferroelectric phase by X-ray diffraction and second-harmonic generation. However, the knowledge of polarization evolution of a single vortex such as nucleation, propagation and stability during phase transition remains largely unknown due to the difficulties in characterizing the small size of the vortices (typically 2–4 nm) in the buried film by the surface probe, X-ray diffraction or optical detection. Hence, a novel route that possesses higher spatial resolution in structural characterization is highly desired.

Herein, we report a method based on local mechanical loading as an alternative stimulus to control vortices by making use of the intrinsic coupling between the strain and polarization[18]. Although previously such a mechanical stress method has been employed to switch the ferroelastic domains in thin films[19, 20], it has never been employed for manipulating the ferroelectric vortices because of the difficulties in fabrication of vortex array and polarization characterization at nanometer scale. We combine atomically resolved *in situ* transmission electron microscopy (TEM) and phase-field modeling methods to study the behavior of the polar vortices in PTO/STO superlattices under mechanical stress. In contrast to conventional substrate-mediated strain control, *in situ* TEM can exert a continuously adjustable stress to a specimen by controlling the indenter freely and allow for real-time observation of the entire switching process, enabling dynamic behavior to be correlated with the applied excitations[20, 21]. In this work, using the custom-built *in situ* TEM holder with much improved stability and carrying out the experiments in a spherical aberration-corrected TEM with sub-angstrom resolution allows us to directly monitor the evolution of each vortex and further map the local lattice and polarization during manipulating it by external excitations.

We find that the polar vortex transforms into the *a*-domain with in-plane polarization under a compressive strain, and spontaneously recovers to the original vortex state after removal of the stress, demonstrating the reversibility of controlling the vortices by mechanical force. Such a phase transition occurs through the inhomogeneous nucleation and growth of newly formed *a*-domains, which is confirmed by the selected-area electron diffraction (SAED) pattern and atomically resolved high-angle annular dark-field scanning transmission electron microscopy (HAADF-STEM) images. Phase-field modeling precisely reproduces such switching events. The single vortex evolution and the stability of vortices are also discussed. These results provide valuable insights into understanding how the mechanical boundary conditions determine the polar states in PTO/STO superlattices. The demonstrated ability to manipulate the ferroelectric vortices at the atomic scale by mechanical stimuli may be valuable for the design of novel electromechanical nanoscale ferroelectric devices.

Thin $(PTO)_n/(STO)_n$ ($n$ denotes the thickness of STO and PTO in terms of the number of unit cells) films were grown on $(110)_o$ $DyScO_3$ substrates with a $SrRuO_3$ (SRO) buffer layer by pulsed-laser deposition (subscript o denotes orthorhombic structure). The experimental details are presented in the methods section. Figure 1a shows a low-magnification atomically resolved HAADF image of a $(PTO)_{11}/(STO)_{11}$ film, where the bright and dark contrast indicates the PTO and STO layers, respectively. The sinusoidal array of in-plane and out-of-plane strain, which is easily distinguished by geometric phase analysis (GPA) of the HAADF images in Fig. 1b,c, is indicative of long-range ordered arrays of the domain structures. Figure 1d shows a polarization map of the PTO layer obtained by calculating the offsets between the Pb and Ti sublattices[22, 23], with the overlaid vectors representing the magnitude and direction of the polar displacements. As a result, the polarization exhibits a continuous rotation, resulting in a vortex structure similar to that reported in a previous study[3]. A SAED pattern acquired from a region that only includes the PTO/STO superlattice layers is shown in Fig. 1e. The equally spaced spots in the out-of-plane direction reflect the long-range order of the superlattices. Close inspection reveals another set of reflections along the in-plane direction similar to that observed from X-ray diffraction pattern[3, 17], indicating long-range ordering of vortices in each layer. The in-plane periodicity of the vortices is ~8.6 nm in Fig. 1f, and the vortex array can also be represented as a displacement map, as shown in Supplementary Fig. 1. Through the SAED, GPA analysis and polarization map based on HAADF images, we demonstrate a long-range order characterized vortices in the as-grown superlattices.

The $(PTO)_{11}/(STO)_{11}$ films were then subjected to *in situ* mechanical stimulation via a scanning probe, as depicted schematically in Fig. 2a. A time series of dark-field images is shown in Fig. 2b, which was selected from the Supplemental Material (Movie S1). Tiny particle-like contrast due to the vortex structure is indicated by the *g*-vector $(200)_{pc}$ excited under two-beam imaging (see Methods; the dark field under the $(002)_{pc}$ vector is shown in Supplementary Fig. 2). The outward progression of the domain area from the nucleation site at the top surface is indicated by the yellow outlines in Fig. 2b, which denote the switching area. The nucleation area is plotted as a function of time and is shown in Fig. 2c, and the change in the area of the domain can be used to roughly estimate the switching velocity. The irregular shape in the transition area and the highly fluctuating switching velocity are likely due to pinning and/or the inhomogeneous distribution of the strain field caused by the irregular geometry of the tungsten tip. After removal of the applied stress, the domain returns to its original state. The mechanical force required to drive vortex transition is estimated to be a few µN, as shown in Fig. 2d,e. Using the Hertz model[24], the maximum stress, $P_{max}$, can be calculated by the formula $P_{max} = 3F/2\pi r^2$, where $F$ is the force and $r$ is the contact radius. Given $r = 50$ nm and $F = 3.14$ µN according to the experimental data Fig. 2e, the resulting stress is approximately 0.6 GPa. This value is indeed lower compared with the stress reported for conventional *a* or c domains switching in $Pb(Zr_{0.2}Ti_{0.8})O_3$[21] or $BaTiO_3$ films[25], indicating the transition of vortices is less energy-consuming and may be feasible for device applications.

We used the phase-field method to model quad-layer $(PTO)_{10}/(STO)_{10}$ superlattices with an STO layer acting as the top layer exposed to air. Supplementary Fig. 3a-c shows the transformation of polar vortices under a mechanical load. The application of mechanical force causes a change in domain structure from polar vortices (Supplementary Fig. 3a) to in-plane polarization (Supplementary Fig. 3b), in excellent agreement with the experimental observations. As expected, larger applied force magnitudes lead to a greater fraction of vortex to in-plane polarization changes (Supplementary Fig. 3c).

The mechanical transition of vortices and the reversibility is further confirmed by the SAED patterns, as shown in Fig. 3a (Movie S2). At 50 s in Fig. 3b, the reflections belonging to the periodic vortex array along the in-plane direction disappear while the reflections of the superlattices along the out-of-plane direction are still visible, indicating the vortex has been switched. Because the intensity is correlated with the number of ordered vortices in the selected area, we plotted the intensity profile of the vortex spots normalized by the superlattice spots (Fig. 3c). This shows a dramatic decrease between 16 and 35 s, and the intensity of superlattice reflections returns to its original value after ~60 s. Hence, it is reasonable to infer that most vortices in the PTO layer has been annihilated under an increasing mechanical stress during 16 s and 35 s in the selected area and recovered immediately after removal of the stress.

The in-plane lattice parameter *a*, out-of-plane lattice parameter *c*, and their ratio (*c*/*a*) as a function of time were calculated from the electron diffraction pattern and plotted in Fig. 3d. Under mechanical stress, the out-of-plane parameter decreases gradually from ~405 pm after 16 s, reaches a minimum plateau of ~ 390 pm after 35 s and returns to its original value at ~60 s as the external stress is removed, whereas the in-plane lattice only slightly increases during load application. Therefore, the ratio of the out-of-plane to in-plane lattice parameter decreases from 1.03 to 0.98, indicating that the newly formed phase is *a*-domain. Since the vortices have different properties in piezoelectric and nonlinear optical response and exhibit exotic electrotoroidic, pizeotoroidic and pyrotoroidic properties arose from the toroidal moment, the formed *a*-domain supposes to have very different properties such as dielectricity[26], piezoelectricity[27] and pyroelectricity[28] than the original vortex. Therefore, our ability to control the transition between vortices and ferroelectric phase allows one to largely tune the responses and susceptibilities.

Single-vortex evolution was captured by the chronological high-resolution TEM image series in Fig. 4a (Movie S3). The inverse fast Fourier transform (IFFT) image of Fig. 4a is shown in Supplementary Fig. 4, through which each vortex can be located. The vortex number decreases with a small-time delay between each layer (Supplementary Fig. 5a), implying that for the transition in each PTO layer to occur there may be a critical stress, which provides the possibility of accurately controlling vortex switching in each layer.

To confirm the atomic structure and polarization after disappearance of the vortex, a series of atomically resolved HAADF images was recorded during mechanical loading, as shown in Fig. 4b, using a double-tilt *in situ* TEM holder (see Methods). The corresponding GPA images clearly show that mechanical loading leads to the disappearance of the vortex. The boundary denoted by the yellow lines in Fig. 4c is sawtooth-shaped. This is further indicative that the switched domain is *a*-domain, because for *a*-domains, the inclined (110)-oriented domain walls have the lowest energy, whereas for *c*-domains, the vertical (100)-oriented domain walls are the most stable ones[29]. From the HAADF image, the polarization vector map can be extracted from the off-center displacements between Pb and Ti as shown in Fig. 4d, which further confirms the polarization is in-plane in the switched area.

Although the vortices can be annihilated and recovered, they are not mobile during mechanical loading, based on the high-resolution TEM images and atomically resolved HAADF images. To determine the stability of the vortex, we selected the third-layer PTO in Fig. 4a, and determined the positions of the vortices by IFFT. The trajectory is plotted in Supplementary Fig. 5b, and the lines at 9.0 and 11.1 s are straight and overlapped. The vortex disappears in the next 150 ms after a total of 11.1 s, with no movement observed in our experiment. Owing to a series of HAADF images in Fig. 4b, the stability of vortices can be further confirmed. The positions of the vortices determined by these GPA images in Supplementary Fig. 6 (see Methods) do not change at all within the uncertainty limits of the experiment. This stability is ideal for data storage because it will reduce the incidence of missing data or cross-talk.

The atomically resolved images during switching indicated that there may be a critical value of stress to switch a vortex. In Supplementary Fig. 7, the vortex stably sustains until at 10.95 s. The switching seems to be initiated in the lower part of the vortex near the tungsten tip and is then quickly pushed forward to the vortex core (11.1 s in Supplementary Fig. 7a). As the core is the most distorted part, it requires a large amount of energy derived from the mechanical stress to overcome the barrier in Landau energy and break the distorted vortex core. Once the core is broken, the entire vortex is quenched quickly within 150 ms. This behavior is different from that proposed in previous theoretical work[30], whereby vortices were predicted to move and melt to reduce the electrostatic energy under an electric field. In contrast, in our case, the compressive out-of-plane stress drives the rotation of polarization toward the in-plane direction, leading to a reduction in the elastic energy. After retraction of the external stress, the vortices are recovered spontaneously because the boundary conditions return to the original state and drive the formation of vortices. Hence, our experiment shows that by modulating the mechanical load, we can control the phase transition process and manipulate the electric toroidal order on the nanometer scale.

Previous seminal work[1] has classified the feasibility of a new device by making use of the toroidal vortices since the electric field inside the vortex is non-uniform due to the extinct polarization configurations[31]. The PTO/STO superlattice systems are superior to data storages because the vortices inside them have a long-range character, which would otherwise require an arranged arrays of nanodots, nanocomposites or nanowires to construct a vortex array. Compared with the unfavorable methods like torque[32], curl electric field[14] or sweeping electric field[33] to induce the transition of the toroid order, the mechanical methods we adopted is readily accessible — only an indenter with small mechanical force is needed. Furthermore, the mechanical force would not cause a big fluctuation in temperature, while the electric current naturally induces more heat in the devices, which would degenerate the vortices to *a* domains when the temperature went up to ~ 473 K[17]. The limitation we have is that now the precise quantitative relation between every single vortex in each PTO layer is absent, disabling the accurate control of every single vortex independently in this arrays in supperlattice. To solve this, a more complicated device with delicate mechanical manipulation is need in future.

In summary, we performed a systematic study of the dynamics of polar vortices in PTO/STO superlattices under a mechanical stimulus using an in situ (S)TEM technique. Our results demonstrate that vortices transform into *a*-domains under mechanical compressive loads and spontaneously return after removal of the external loads at the atomic level. These results provide valuable insights into understanding the fundamental properties of the topological defects in ferroics. The demonstrated ability to control single vortex by external stress paves the way toward the development of advanced high-density storage devices.

**References**
1. Naumov, I. I. *et al.* Unusual phase transitions in ferroelectric nanodisks and nanorods. *Nature*. **432**, 737-740 (2004).
2. Jia, C. L. *et al.* Direct observation of continuous electric dipole rotation in flux-closure domains in ferroelectric Pb(Zr,Ti)$O_3$. *Science*. **331**, 1420-1423 (2011).
3. Yadav, A. K. *et al.* Observation of polar vortices in oxide superlattices. *Nature*. **530**, 198-201 (2016).
4. Shafer, P. *et al.* Emergent chirality in the electric polarization texture of titanate superlattices. *PNAS*. **115**, 915-920 (2018).
5. Pilania, G. & Ramprasad, R. Complex polarization ordering in PbTi$O_3$ nanowires: A first-principles computational study. *Phys. Rev. B*. **82**, 155442 (2010).
6. Prosandeev, S. *et al.* Controlling toroidal moment by means of an inhomogeneous static field: an ab initio study. *Phys. Rev. Lett*. **96**, 237601 (2006).
7. Nahas, Y. *et al.* Frustration and self-ordering of topological defects in ferroelectrics. *Phys. Rev. Lett*. **116**, 117603 (2016).
8. Kim, K. E. *et al.* Configurable topological textures in strain graded ferroelectric nanoplates. *Nat. Commun.* **9**, 403 (2018).
9. Balke, N. *et al.* Enhanced electric conductivity at ferroelectric vortex cores in BiFe$O_3$. *Nat. Phys.* **8**, 81-88 (2011).
10. Nelson, C. T. *et al.* Spontaneous vortex nanodomain arrays at ferroelectric heterointerfaces. *Nano Lett.* **11**, 828-834 (2011).
11. Mei, A. B. *et al.* Discovery of ordered vortex phase in multiferroic oxide superlattices. *arXiv:1810.12895v1*; 2018.
12. Rodriguez, B. J. *et al.* Vortex polarization states in nanoscale ferroelectric arrays. *Nano Lett.* **9**, 1127-1131 (2009).
13. Scott, J. F. Applications of modern ferroelectrics. *Science*. **315**, 954-959 (2007).
14. Wang, J. & Kamlah, M. Intrinsic switching of polarization vortex in ferroelectric nanotubes. *Phys. Rev. B*. **80**, 012101 (2009).

15. Prosandeev, S. *et al.* Control of vortices by homogeneous fields in asymmetric ferroelectric and ferromagnetic rings. *Phys. Rev. Lett*. **100**, 047201 (2008).
16. Van Lich, L. *et al.* Switching the chirality of a ferroelectric vortex in designed nanostructures by a homogeneous electric field. *Phys. Rev. B*. **96**, 134119 (2017).
17. Damodaran, A. R. *et al.* Phase coexistence and electric-field control of toroidal order in oxide superlattices. *Nat. Mater*. **16**, 1003-1009 (2017).
18. Chen, W. J. *et al.* Vortex domain structure in ferroelectric nanoplatelets and control of its transformation by mechanical load. *Sci. Rep*. **2**, 796 (2012).
19. Sun, C. T. & Achuthan, A. Domain-switching criteria for ferroelectric materials subjected to electrical and mechanical loads. *JACS*. **87**, 395-400 (2004).
20. Chen, Z. *et al.* Facilitation of ferroelectric switching via mechanical manipulation of hierarchical nanoscale domain structures. *Phys. Rev. Lett*. **118**, 017601 (2017).
21. Gao, P. *et al.* Ferroelastic domain switching dynamics under electrical and mechanical excitations. *Nat. Commun*. **5**, 3801 (2014).
22. Tang, Y. L. *et al.* Observation of a periodic array of flux-closure quadrants in strained ferroelectric $PbTiO_3$ films. *Science*. **348**, 547-551 (2015).
23. Lu, L. *et al.* Topological defects with distinct dipole configurations in $PbTiO_3/SrTiO_3$ multilayer Films. *Phys. Rev. Lett*. **120**, 177601 (2018).
24. Gaillard, Y. *et al.* Nanoindentation of $BaTiO_3$: dislocation nucleation and mechanical twinning. *J. Phys. D: Appl. Phys*. **42**, 085502 (2009).
25. Lu, H. *et al.* Mechanical Writing of Ferroelectric Polarization. *Science*. **336**, 59-61 (2012).
26. Karthik, J. *et al.* Effect of 90 degrees domain walls on the low-field permittivity of $PbZr_{0.2}Ti_{0.8}O_3$ thin films. *Phys. Rev. Lett*. **108**, 167601 (2012).
27. Zednik, R. J. *et al.* Mobile ferroelastic domain walls in nanocrystalline PZT Films: the direct piezoelectric effect. *Adv. Funct. Mater*. **21**, 3104-3110 (2011).
28. Karthik, J. *et al.* Effect of 90 degrees domain walls and thermal expansion mismatch on the pyroelectric properties of epitaxial $PbZr_{0.2}Ti_{0.8}O_3$ thin films. *Phys. Rev. Lett*. **109**, 257602 (2012).
29. Meyer, B. & Vanderbilt, D. Ab initio study of ferroelectric domain walls in $PbTiO_3$. *Phys. Rev. B*. **65**, 104111 (2002).
30. Hong, Z. & Chen, L.-Q. Blowing polar skyrmion bubbles in oxide superlattices. *Acta Mater*. **152**, 155-161 (2018).
31. Yadav, A. K. *et al.* Spatially resolved steady-state negative capacitance. *Nature*. **565**, 468-471 (2019).
32. Liu, J. *et al.* Controlling polar-toroidal multi-order states in twisted ferroelectric nanowires. *npj Computational Materials*. **4**, 1-8 (2018).
33. Ma, L. L. *et al.* Direct electrical switching of ferroelectric vortices by a sweeping biased tip. *Acta Mater*. **158**, 23-37 (2018).

## Acknowledgments

This work was supported by the National Equipment Program of China (ZDYZ2015-1), National Key R&D Program of China (2016YFA0300804, 2016YFA0300903), National Natural Science Foundation of China (51672007, 51572233, 61574121, 21773303 and 51421002), and the National Program for Thousand Young Talents of China. The work at Penn State is supported by the U.S. Department of Energy, Office of Basic Energy Sciences, Division of Materials Science and Engineering under Award FG02-07ER46417 (J.A.Z. and L.Q.C.) and partially by a graduate fellowship from the 3M Company (J.A.Z).

**Author contributions**

P.G., C.B.T. and X.D.B. conceived the idea and directed the project. P.C. analyze the data and wrote the paper. P.C. and M.Q.L. performed the experiments assisted by Y.W.S., X.M.L. and Y.H.L. under the direction of P.G. and X.D.B. A.Y.A., X.M.M. and K.H.L. assisted the data analysis. Z.X. provided support for *in situ* TEM holders. C. B.T. and X.L.Z. grew the samples assisted by C.L.R. and J.B.W. J.A.Z. carried out the phase-field simulation supervised by L.Q.C. the All authors discussed the results and commented on the manuscript.

**Competing interests**

The authors declare that they have no competing interests.

**Data availability**

The data that support the plots within this paper and other findings of this study are available from the corresponding author upon reasonable request.

**Figures**

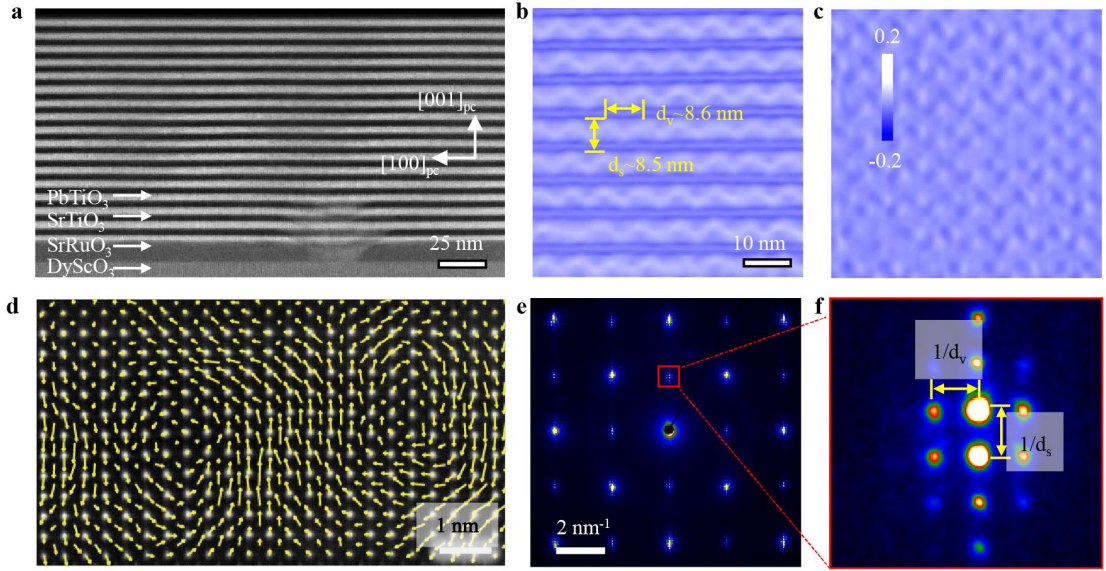

**Fig.1 | Characterization of vortices in PbTiO₃/SrTiO₃ superlattices**. **a**, Low-magnification scanning transmission electron microscopy (STEM) image of a $(PbTiO_3)_{11}/(SrTiO_3)_{11}$ superlattice along $[010]_{pc}$, showing the alternative arrangement of SrTiO₃ and PbTiO₃ on a DyScO₃ substrate. **b,c**, Geometric phase analysis of the STEM data showing the distribution of the in-plane strain $\mathcal{E}_{xx}$ and out-of-plane strain $\mathcal{E}_{yy}$, respectively. **d**, Cross-sectional high-angle annular dark-field (HAADF) STEM image with an overlay of the polar displacement vectors denoted by the yellow arrows showing the vortices in the PbTiO₃ layer. **e**, A selected-area electron diffraction (SAED) pattern for the PbTiO₃/ SrTiO₃ film. **f**, Enlarged (001) spots showing the satellite diffraction spots from the ordered vortex. The vortex and superlattice periods are measured to be 8.6 and 8.5 nm, respectively, denoted by $d_v$ and $d_s$.

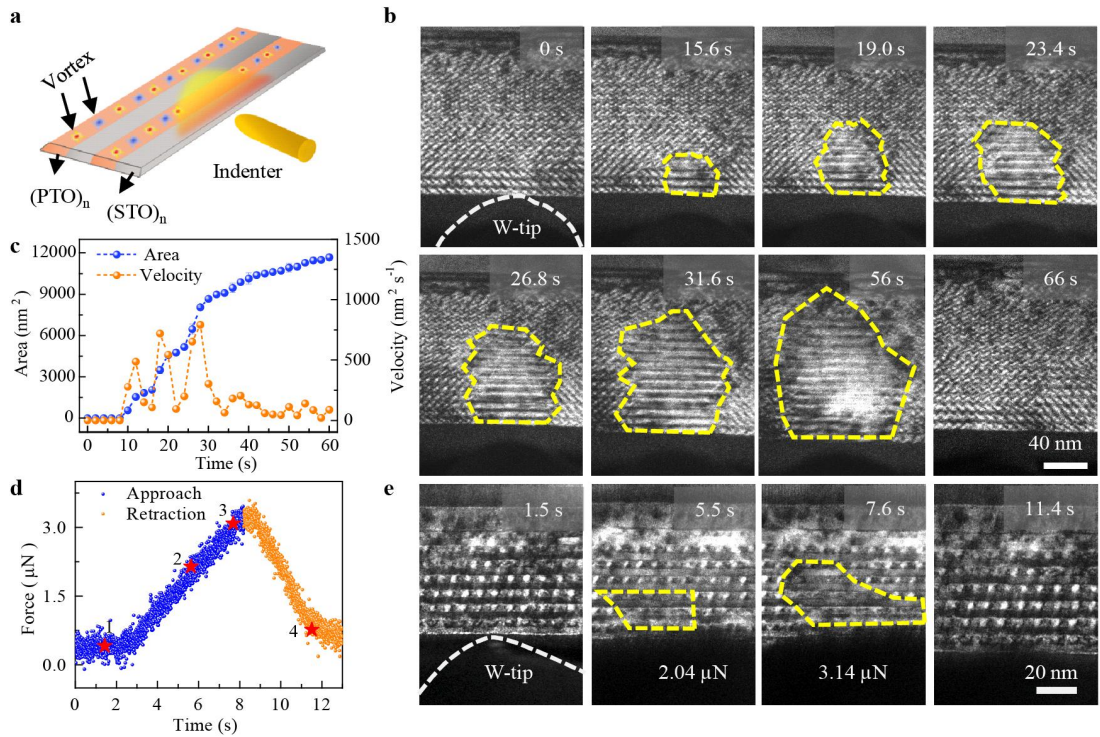

**Fig. 2 | Mechanical manipulation of vortices by *in situ* TEM. a**, Schematic image of the experimental setup, with a mobile tungsten tip acting as an indenter for the mechanical manipulation of vortices. **b**, Chronological TEM dark-field image series formed by reflection with $g = 200_{pc}$. Under the mechanical loads, vortices contrast gradually disappears. **c**, Corresponding transition area (blue line) and switching velocity (orange line) plotted as functions of time. **d**, Mechanical load as a function of time, with the blue points representing the approach branch and orange points corresponding to the retraction branch. The highlighted red stars along with label 1-4 correspond to images in (**e**). **e**, Dark-field images showing the vortex evolution under certain measured mechanical stress.

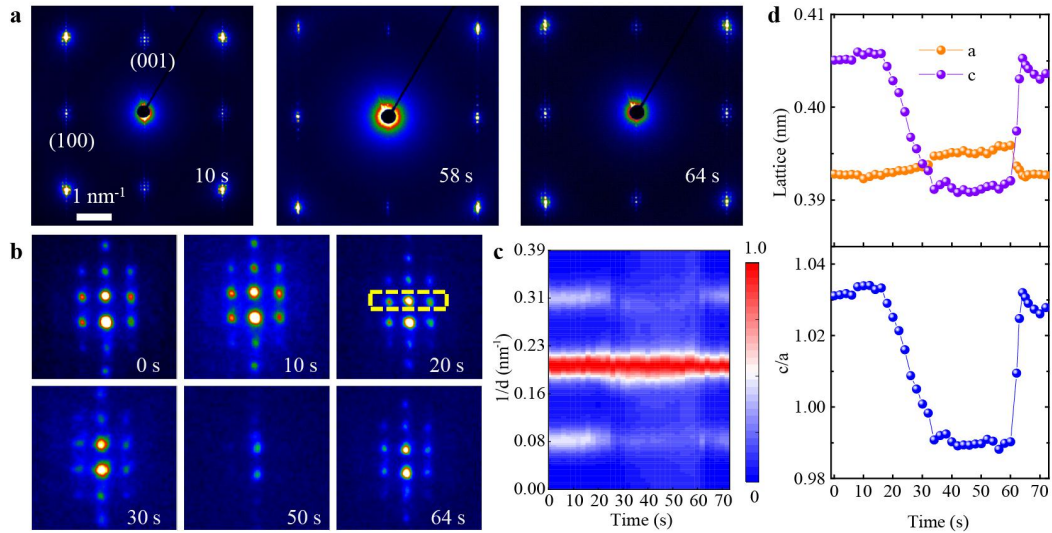

**Fig. 3 | Structural evolution of vortices under mechanical stress. a**, Three SAED images extracted from a real-time image series corresponding to before, during, and after mechanical loading (from left to right). The vortex reflections disappear at 58 s and recover when the external stress is removed at 64 s. **b**, Chronological SAED $(001)_{pc}$ images with vortex spots dimming as the mechanical force is continuously applied, yellow box at 20 s indicates diffraction spots in (**c**). **c**, Line profile intensity as a function of time normalized by the center superlattice spots. **d**, The in-plane *a* and out-of-plane *c* lattice parameters and *c*/*a* ratio as functions of time. The *c*/*a* ratio is eventually less than 1 under a continuously applied mechanical loads, indicating that the vortices have transformed to *a*-domains.

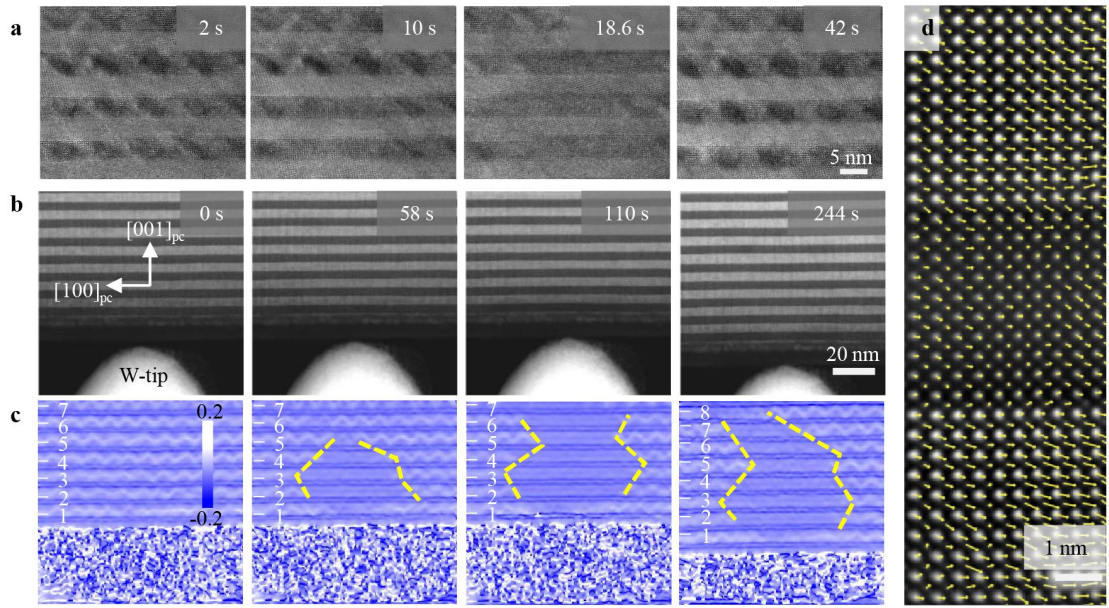

**Fig. 4 | Atomic-scale tracking of the vortex transition. a**, Time-lapse high-resolution TEM images showing the switching process. **b,c,** HAADF-STEM image series acquired under a mechanical load and the corresponding GPA images, respectively. The yellow lines in (**c**) show the boundary between the transformed and untransformed areas. The PTO layer is denoted by the white numbers on the left side along each image. **d**, Atomically resolved STEM image overlaid with displacement arrows showing that the transition area becomes *a*-domains.

18